\documentclass{aa}
\usepackage{graphicx}
\usepackage[varg]{txfonts}
\usepackage{amssymb}
\usepackage{color}
\usepackage[switch, modulo]{lineno} \modulolinenumbers[10] \linenumbers 

\usepackage{natbib}
\usepackage{twoopt}
\usepackage[breaklinks=true]{hyperref} 
\makeatletter
  \newcommandtwoopt{\citeads}[3][][]{\href{http://adsabs.harvard.edu/abs/#3}%
    {\def\hyper@linkstart##1##2{}%
     \let\hyper@linkend\@empty\citealp[#1][#2]{#3}}}
  \newcommandtwoopt{\citepads}[3][][]{\href{http://adsabs.harvard.edu/abs/#3}%
    {\def\hyper@linkstart##1##2{}%
     \let\hyper@linkend\@empty\citep[#1][#2]{#3}}}
  \newcommandtwoopt{\citetads}[3][][]{\href{http://adsabs.harvard.edu/abs/#3}%
    {\def\hyper@linkstart##1##2{}%
     \let\hyper@linkend\@empty\citet[#1][#2]{#3}}}
  \newcommandtwoopt{\citeyearads}[3][][]%
    {\href{http://adsabs.harvard.edu/abs/#3}
    {\def\hyper@linkstart##1##2{}%
     \let\hyper@linkend\@empty\citeyear[#1][#2]{#3}}}
\makeatother
\begin{document} 

   \title{Active region fine structure observed at 0.08$\arcsec$ resolution}

        \author{R.~Schlichenmaier\inst{1} \and O.~von der L\"uhe \inst{1}  \and S.~Hoch\inst{1}
         \and  D.~Soltau\inst{1}\and T.~Berkefeld\inst{1}  \and D.~Schmidt\inst{1,8}  \and W.~Schmidt\inst{1}
         \and C.~Denker\inst{2}   \and H.~Balthasar\inst{2} \and A.~Hofmann\inst{2}  
         \and  K.~G.~Strassmeier\inst{2} \and  J. Staude\inst{2} 
         \and A.~Feller\inst{3}  \and A.~Lagg\inst{3}   \and S.K.~Solanki\inst{3,9} 
         \and M.~Collados\inst{4,5} 
         \and M.~Sigwarth\inst{1} \and R. Volkmer\inst{1}  \and T. Waldmann\inst{1} 
         \and F. Kneer\inst{6} \and H. Nicklas\inst{6} 
         \and M.~Sobotka\inst{7} 
                    }
   \institute{ Kiepenheuer-Institut f\"ur Sonnenphysik,  Sch\"oneckstr. 6, 79104 Freiburg, Germany
   \and Leibniz Institut f\"ur Astrophysik Potsdam (AIP), An der Sternwarte 16, 14482 Potsdam, Germany
   \and Max-Planck-Institut f\"ur Sonnensystemforschung, Justus-von-Liebig-Weg 3, 37077, G\"ottingen, Germany.
   \and Instituto de Astrof\'isica de Canarias (IAC), V\'ia Lact\'ea, 38200 La Laguna (Tenerife), Spain
   \and Departamento de Astrof\'isica, Universidad de La Laguna, 38205 La Laguna (Tenerife), Spain 
   \and Institut f\"ur Astrophysik, Friedrich-Hund-Platz 1, 37077, G\"ottingen, Germany
   \and Astronomical Institute, Academy of Sciences of the Czech Republic, Fri\v{c}ova 298, 25165 Ond\v{r}ejov, Czech Republic 
   \and National Solar Observatory, 3010 Coronal Loop, Sunspot, NM 88349, USA
   \and School of Space Research, Kyung Hee University, Yongin, Gyeonggi 446-701, Republic of Korea
   }

  \date{Received 21/03/2016; accepted 25/07/2016}
  
 \abstract
 {The various mechanisms of magneto-convective energy transport determine the structure of sunspots and active regions. }
{We characterise the appearance of light bridges and other fine-structure details and elaborate on their magneto-convective nature.}
{We present speckle-reconstructed images taken with the broad-band imager (BBI) at the 1.5 m GREGOR telescope in the 486\,nm and 589\,nm bands. We estimate the spatial resolution from the noise characteristics of the image bursts and obtain 0.08$\arcsec$ at 589\,nm. We describe structure details in individual best images as well as the temporal evolution of selected features.}
{We find branched dark lanes extending along thin ($\approx\!1\arcsec$) light bridges in sunspots at various heliocentric angles. In thick ($\gtrsim\!2 \arcsec$) light bridges the branches are disconnected from the central lane and have a Y shape with a bright grain toward the umbra. The images reveal that light bridges exist on varying intensity levels and that their small-scale features evolve on timescales of minutes. Faint light bridges show dark lanes outlined by the surrounding bright features. Dark lanes are very common and are also found in the boundary of pores. They have a characteristic width of 0.1$\arcsec$ or smaller. Intergranular dark lanes of that width are seen in active region granulation.}
{We interpret our images in the context of magneto-convective simulations and findings: While central dark lanes in thin light bridges are elevated and associated with a density increase above upflows, the dark lane branches correspond to locations of downflows and are depressed relative to the adjacent bright plasma. Thick light bridges with central dark lanes show no projection effect. They have a flat elevated plateau that falls off steeply at the umbral boundary. There, Y-shaped filaments form as they do in the inner penumbra. This indicates the presence of inclined magnetic fields, meaning that the umbral magnetic field is wrapped around the convective light bridge.}
%
%
\maketitle

\section{Introduction}

Sunspots are the largest manifestation of magnetic fields in the solar photosphere and appear darker than the surrounding photosphere in which convective heat transport takes place in the form of granulation \citep{solanki2003,schlichenmaier2009}. The details of the heat transport within sunspots are still under debate. It is clear, however, that it is of convective origin: \citet{deinzer1965} concluded that convective transport of heat must take place even in the coolest parts of sunspots, since neither heat conduction nor radiative heat transport suffices to sustain the observed umbral temperatures. 

The appearance of sunspots is characterised by different modes of radiatively driven magneto-convection \citep[cf.][]{2011LRSP....8....3R}: In sunspot penumbrae, magneto-convection takes place in inclined magnetic fields, where it generates elongated bright and dark filaments \citep{borrero+ichimoto2011}. In the cooler umbra, the magnetic field is predominantly vertical and strong, suppressing vigorous convection. \citet{schuessler+voegler2006} suggested that radiative losses drive small magneto-convective cells, which are observed as umbral dots  \citep{2010A&A...510A..12B}. In these cells, hot rising plumes cause the umbral dots to appear bright. The vertical magnetic field is diminished by (a) adiabatic expansion of the rising hot bubble and (b) diverging flows that advect the vertical field lines out of the cell. These new insights replaced older concepts, according to which convective columns of unmagnetised plasma intrude into the umbra and "punch their way up" \citep{parker1979} to appear as field-free umbral dots in the photosphere. According to the numerical models, umbral dots are associated with upflows and have a central dark lane in intensity images. \citet{rimmele+marino2006} and \cite{rimmele2008} confirmed these characteristics with observations at high spatial resolution.


\begin{figure*}
\centering
\includegraphics*[width=15cm]{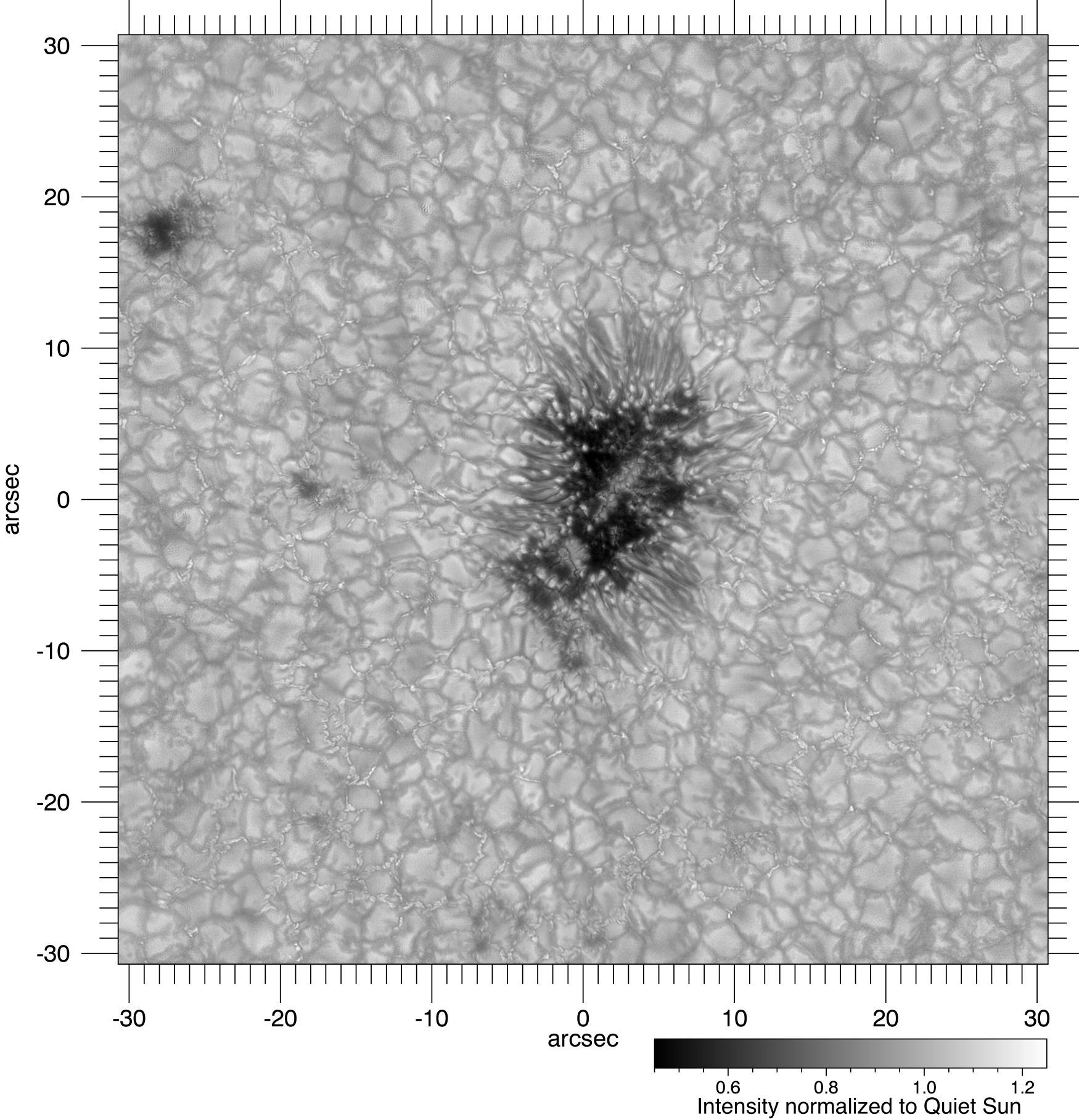}
\caption{\label{fig:1} 
Spot of NOAA 11757, observed on 31 May 2013, 12:35 UT at 589\,nm.  
It was in its decaying phase at the time of our observation and disappeared from the disc on 3 June 2013.
}
\end{figure*}

While it is generally accepted that umbral dots always occur in sunspots and it is only a matter of resolving them \citep[e.g.][]{riethmueller+al2008, kilcik+al2012, 2014PASJ...66S...1W}, the phenomenon of light bridges is not seen in every spot.  \citet{sobotka1997} reviewed the structure of light bridges and classified them depending on whether they separate the umbral core into strong and faint.
Another criterion is their internal structure, which can be granular or filamentary. Light bridges can be associated with chromospheric brightenings \citep{2003ApJ...589L.117B}. \citet{2006A&A...453.1079J} suggested that these brightenings are a result of vigorous magneto-convective motions in the deep layers of a light bridge. By those motions, magnetic field lines are advected along the light bridge, creating magnetic tension. This tension propagates into the chromosphere and is released there. This analysis, based on depth-dependent inversions of Stokes profiles, revealed that the magnetic field lines meet above the light bridge. The authors found evidence that the point of closure is deeper in the atmosphere for smaller light bridges. \citet{2014A&A...568A..60L} found strong convective motions within a large granular light bridge and confirmed the topological picture proposed by \citet{2006A&A...453.1079J}. 

Insights into the physical processes can be gained from the morphological intensity structure at very high spatial resolution. \citet{2003ApJ...589L.117B} and \citet{lites+al2004} observed filamentary light bridges with a central dark lane at an angular resolution of 0\farcs12. Elongated granules were observed on both sides of this central lane, which
were intercepted by dark lines perpendicular to the central dark lane. In this contribution, we study the brightness morphology of light bridges at a resolution of some 0.08$\arcsec$, that
is, at the smallest currently resolvable spatial scales. We present images with unprecedented resolution and elaborate on the morphological characteristics of light bridges, their dark lanes, and other fine-structure details.

\begin{table}[b]
\parindent 0em
\caption{\label{table_1}Data description.}
\begin{tabular}{lcccccc}
2013      & UT  & NOAA & info & Position & $\theta$ & Fig. \\\hline\hline \rule[-3pt]{0pt}{12pt}
31 May  & 12:34 & 11757 & spot & S08W14 & 9   & \ref{fig:3r} \\ \hline  \rule[-3pt]{0pt}{12pt}
9 June   & 09:22 & 11765 & pore & N08W50 & 26 & \ref{fig:4r} \\ \hline  \rule[-3pt]{0pt}{12pt}
13 June & 09:16 & 11768 & p-sp. & S11W33 & 25 &  \ref{fig:1r} \\ \hline  \rule[-3pt]{0pt}{12pt}
14 June & 08:26 & 11768 & f-sp. & S11W60 & 37 &  \ref{fig:2ra} \\ \hline  \rule[-3pt]{0pt}{12pt}
14 June & 09:23 & 11768 & spot & S11W60 & 41 &  \ref{fig:2rb} \\ \hline 
\end{tabular}\\
$\theta$: heliocentric angle [deg]; p: preceding; f: following; sp.: spot 
\end{table}

\section{Observation, data calibration, and reduction}

\subsection{Observations and preprocessing}

We observed active regions, mostly sunspots and pores, with the broad-band imager \citep[BBI,][]{2012AN....333..894V} of the 1.5\,m GREGOR telescope \citep{2012AN....333..796S, 2012AN....333..847S} at the Observatorio del Teide in Tenerife, Spain. The observations presented in this publication were carried out during the early-science phase in May and June 2013. 
Table~\ref{table_1} presents the AR numbers, heliographic positions, and the dates and times of these observations. 

We used both imaging channels of the BBI and observed the continuum restricted to spectral bands at 430\,nm (FWHM 8 nm), 486\,nm (FWHM 4\,nm) and/or 589\,nm (FWHM 19\,nm) using interference filters. Here we present data taken on 31 May 2013 with a PCO 4000 camera with 4008 by 2672 pixels at 589\,nm with 0\farcs030/px. All other presented data sets were taken with a PCO 4000 at 486\,nm with 0\farcs030/px.
Seeing and static aberrations were partially compensated for
by the GREGOR adaptive optices system \citep[GAOS,][]{2012AN....333..863B}.
We simultaneously acquired bursts of images with an exposure of 1\,ms, which is short enough to freeze the smearing effects of the seeing.  To achieve a high frame rate, the detector was read out in parallel by two analog digital unit converters. A burst of 100 frames could be collected within 20 seconds and stored in camera local memory with the PCO 4000, which is just short enough for sampling the time evolution in the photosphere at small scales. Acquisition, conversion, transfer to the control computer and writing to hard disc of a burst takes altogether 90 seconds, setting the time resolution limit for studying dynamic phenomena on the Sun.

The images were prepared for speckle image reconstruction using a data pipeline implemented in IDL. 
First the image burst is corrected for dark current and gain table.
Flatfield data are regularly acquired by pointing the telescope rapidly along a circular path around the solar disc centre with an adjustable radius such as to avoid sunspots and active regions. 
The two-channel readout of the camera introduces a mismatch of mean intensities between the two sensor halves that is not fully removed by the flatfielding procedure. 
We calculate the spatial means over the respective areas of the flat-field corrected images to appropriately scale the sensor halves. 
Residual flat-field artefacts are produced by dust in the beam close to a focal plane and by the slowly evolving misalignment of the cameras. 
The procedure for their removal was described in detail by \citet{vdluehe1+1993}. 
After dark and flatfield correction, we retain the best images of the burst (typically 80\%) based on intensity contrast. 
This procedure discards images that are blurred due to poor seeing in conjunction with loss of lock of the AO compensation. 

\begin{figure}
\includegraphics*[width=9cm]{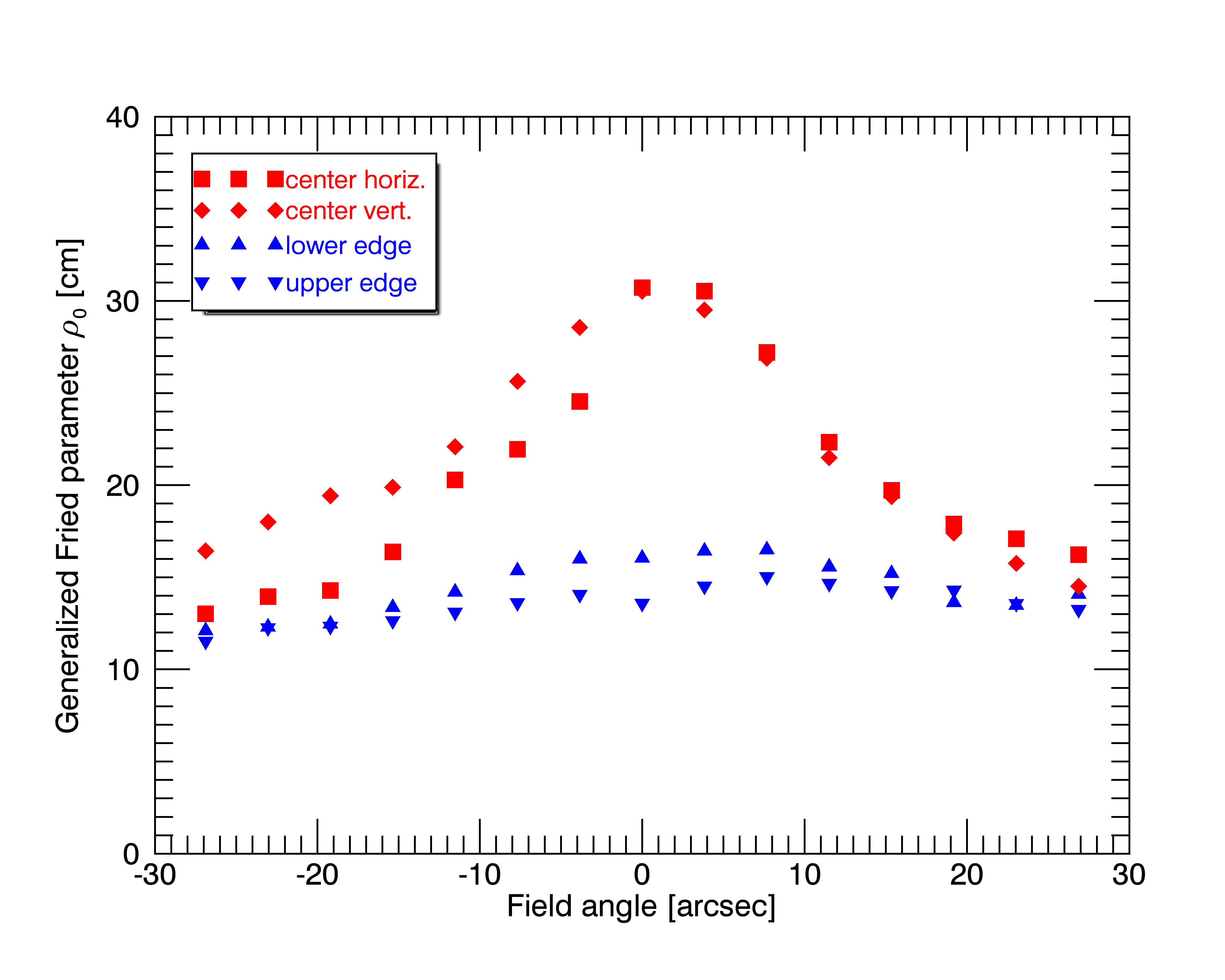}
\caption{\label{fig:2} 
Generalised Fried parameter $\rho_0$ across the field shown in Fig.~\ref{fig:1}. 
Each data point represents a reconstructed subfield.
Squares and diamonds cross the centre of the field and show the increase at the GAOS lock point, while triangles show the background value near the edge of the field.
}
\end{figure}

\begin{figure}
\includegraphics*[width=9cm]{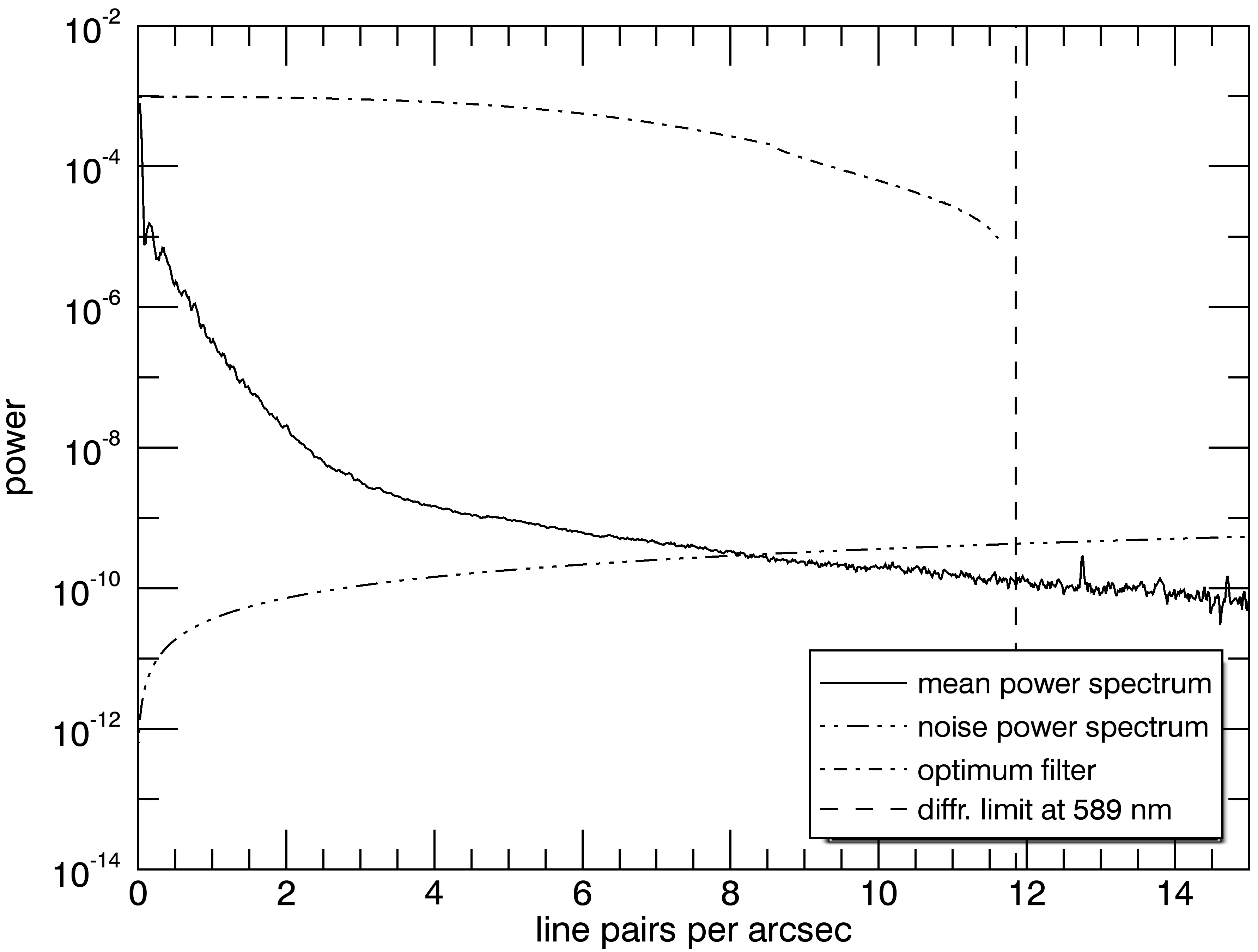}
\caption{\label{fig:3} 
Mean power spectrum representing the full field corrected for noise bias, estimated noise in the spectrum, and optimum filter (multiplied by $10^{-3}$).
}
\end{figure}

\begin{figure}
\includegraphics*[width=9cm]{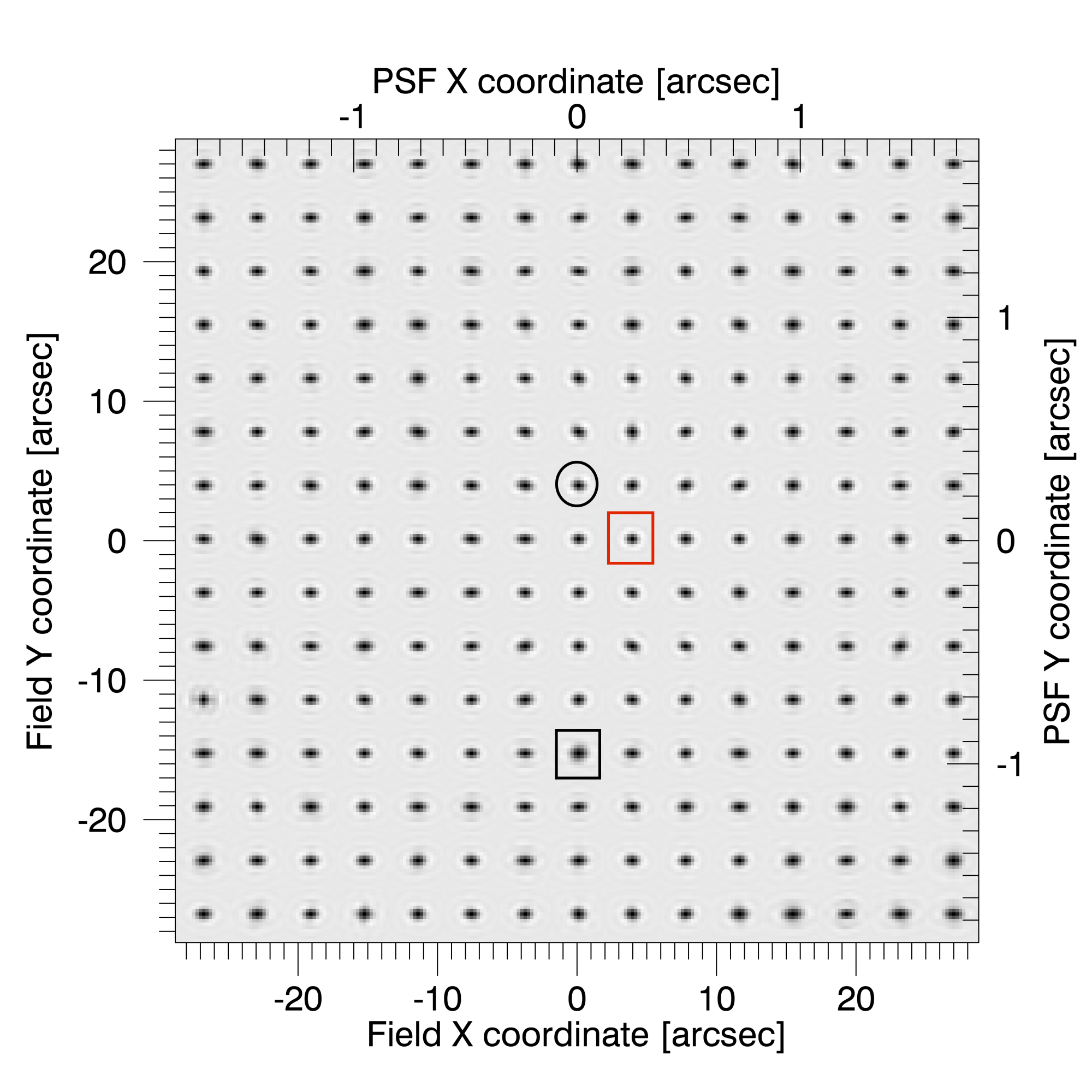}
\caption{\label{fig:4} 
Point spread functions derived from the noise filter centred at the positions of their subfields.
The PSFs are shown magnified in scale with inverted intensity to emphasise asymmetry and ringing. 
The bottom and left axes represent the full field coordinates of the position of each PSF, the top and right axes show the dimension of each PSF.
The GAOS lock point is marked by the black circle. The red and black squares correspond to Fig.~\ref{fig:5}. 
}
\end{figure}

\begin{figure}
\includegraphics*[width=9cm]{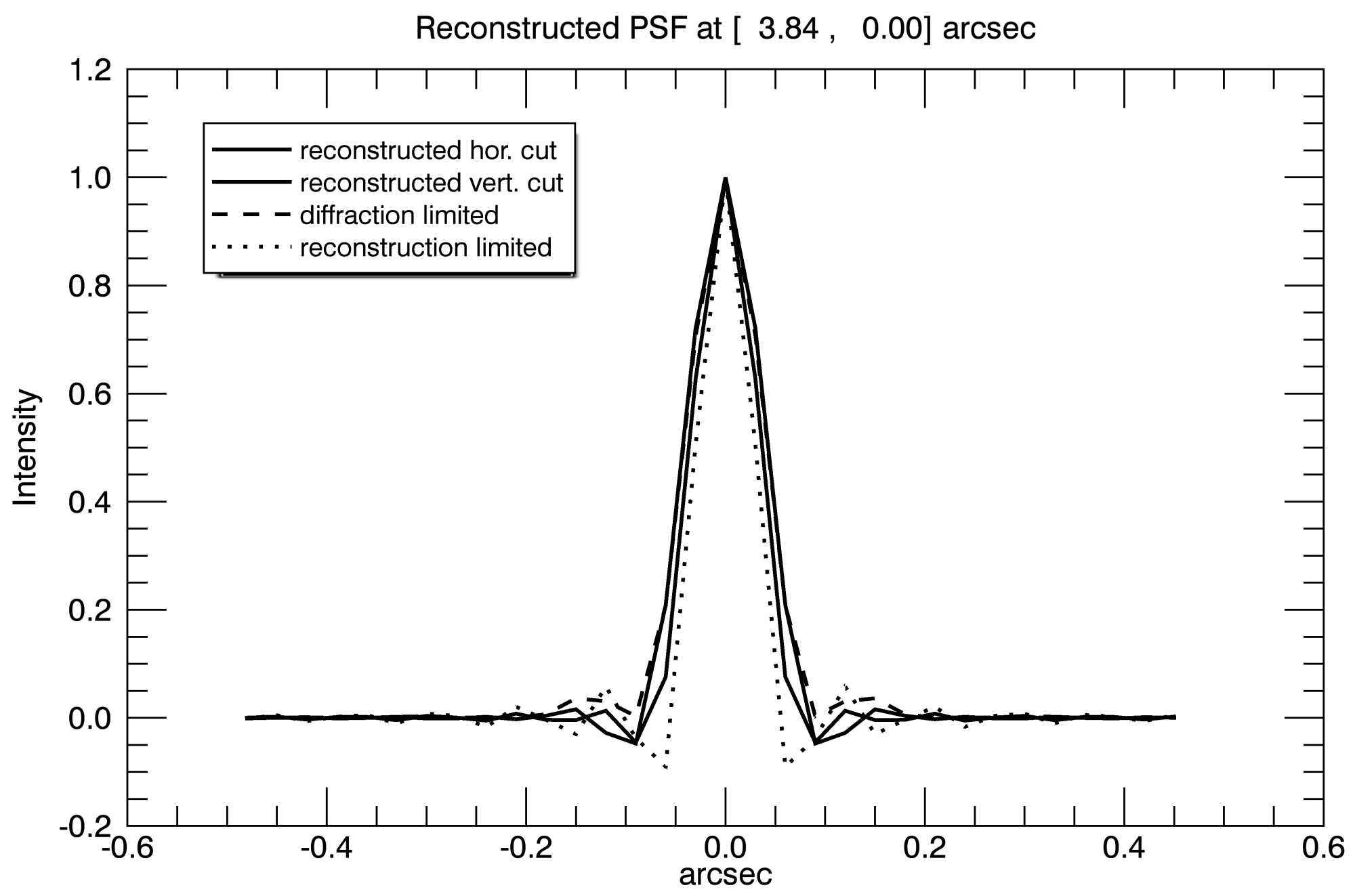}
\includegraphics*[width=9cm]{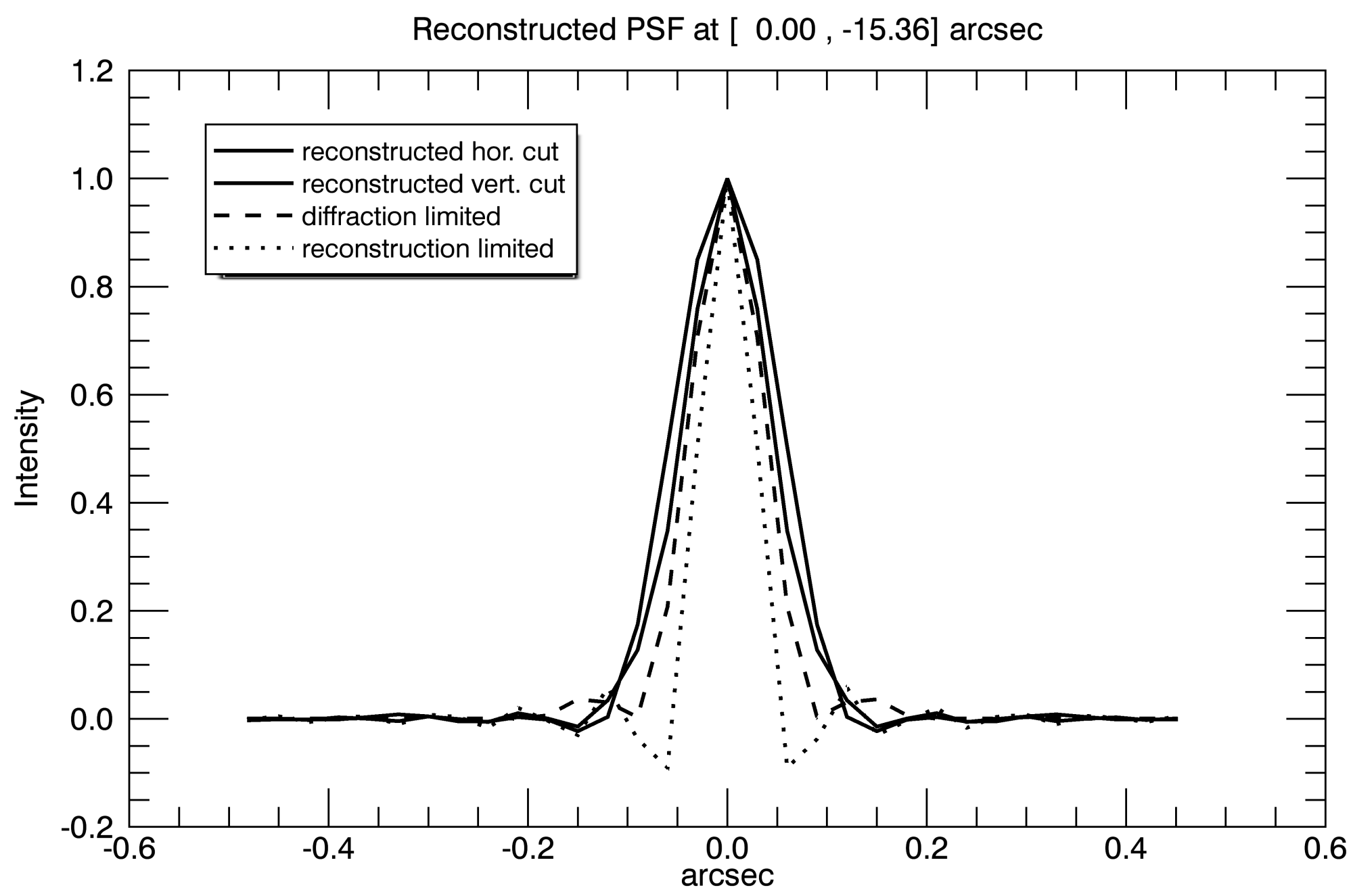}
\caption{\label{fig:5} 
Best (upper panel, corresponding to the red square in Fig.~\ref{fig:4}) and poorest (lower panel, corresponding to the black square in Fig.~\ref{fig:4}) noise filter point spread functions (solid lines) compared with the diffraction-limited PSF of GREGOR at 589 nm (dashed) and a hypothetical reconstruction-limited noise filter PSF for which all signal within the diffraction limit is recovered (dotted lines). 
}
\end{figure}

\subsection{Image reconstruction, resolution, and photometric fidelity}

We analysed a data set taken on 31 May 2013 at 12:35 UT to investigate the performance of the telescope, instruments, and data analysis procedures.
This set consists of a 100 frame burst, which was taken within 20 seconds.
We selected a region of 2048 by 2048 pixels, corresponding to about 1 x 1 arcmin and centred on a small sunspot for further analysis. 
Speckle imaging requires dissection of the field of view into overlapping subfields to account for anisoplanatism \citep{vdluehe1+1993}.
We divided the burst into 15 x 15 sub-fields with a side length of 7\farcs7, overlapping by 50\% of their width in either direction.
Each sub-field was individually reconstructed, and the results were reassembled to yield the image in Fig \ref{fig:1}.
Each subfield reconstruction comprises Fourier amplitude estimation using the Labeyrie \citep{labeyrie+1970} and spectral ratio \citep{vdluehe+1983}  methods and Fourier phase estimation using the Knox--Thompson \citep{mikurda+2006} and triple correlation \citep{woeger+vdluhe2008} algorithms.  

The wavefront sensor of GAOS uses a $12\arcsec$ field centred on the sunspot as the lock point for compensation.
The compensation deteriorates somewhat with increasing field angle from the lock point, leading to a variation of the generalised Fried parameter $\rho_0$ \citep{cagigal+2000}, Fig.~\ref{fig:2}.
In principle, speckle imaging recovers Fourier amplitudes of the solar structure up to the diffraction limit, with the effects of seeing and the modulation transfer function of the telescope removed.
In practice, the achieved resolution depends on the signal-to-noise ratio of the reconstructed Fourier amplitude, which in turn depends on $\rho_0$ and on the amount of small-scale structure present in the reconstructed scene.
A good measure of the actually achieved spatial resolution is the noise filter, which is constructed from the uncompensated Fourier amplitudes and the noise spectrum \citep[see ][]{vdluehe1+1993}, Fig.~\ref{fig:3}, and represents the transfer function of the overall reconstruction process.
Ideally, this function would be unity up to the diffraction limit and vanish beyond.
In practice, noise filters decrease to zero somewhere between 50\% and 90\% of the diffraction limit.
The Fourier transform of the noise filter therefore represents the effective point spread function of the reconstruction process. 
Each subfield reconstruction uses an individual noise filter, depending on the distance from the lock point and the complexity of the scene.
A matrix of reconstruction point spread functions across the full field is shown in Fig.~\ref{fig:4}. Cross sections of a best and worst case are compared with the reconstruction-limited PSF (Fourier transform of a filled circle with the radius of the diffraction limit) and the diffraction-limited PSF of the telescope (including central obstruction) and are shown in Fig.~\ref{fig:5}.
The PSF at the lock point close to the centre of the field is indistinguishable from a diffraction-limited PSF and has a FWHM of 0.08 arcsec, while the only slightly broader worst-case PSF has a FWHM of 0\farcs1.

It is remarkable to observe that although the compensation by GAOS varies significantly across the field, the PSFs of the reconstruction process do not.
The performance of the reconstruction depends mostly on the small-scale structure content of the scene for the degree of compensation achieved here.
Many PSFs are elongated, which is partially caused by anisotropy of the observed scene, but may also be caused by anisotropic statistics of partially compensated seeing or by instrumental effects.

\citet{woeger+al2008} demonstrated by comparison with observations from space that speckle imaging provides reconstructed images with high photometric fidelity. We compared image reconstructions from two bursts that are taken almost simultaneously in both imaging channels with broad-band filters at 486 and 430\,nm. We verified that all the small-scale structure in both images is morphologically identical. If any of the small-scale structure were due to reconstruction artefacts, they would only be seen in one of the images. For all observed features that we discuss in this paper, we checked that they are visible in both image channels. We only show the image with the better quality.

\begin{figure}
\includegraphics[width=9cm]{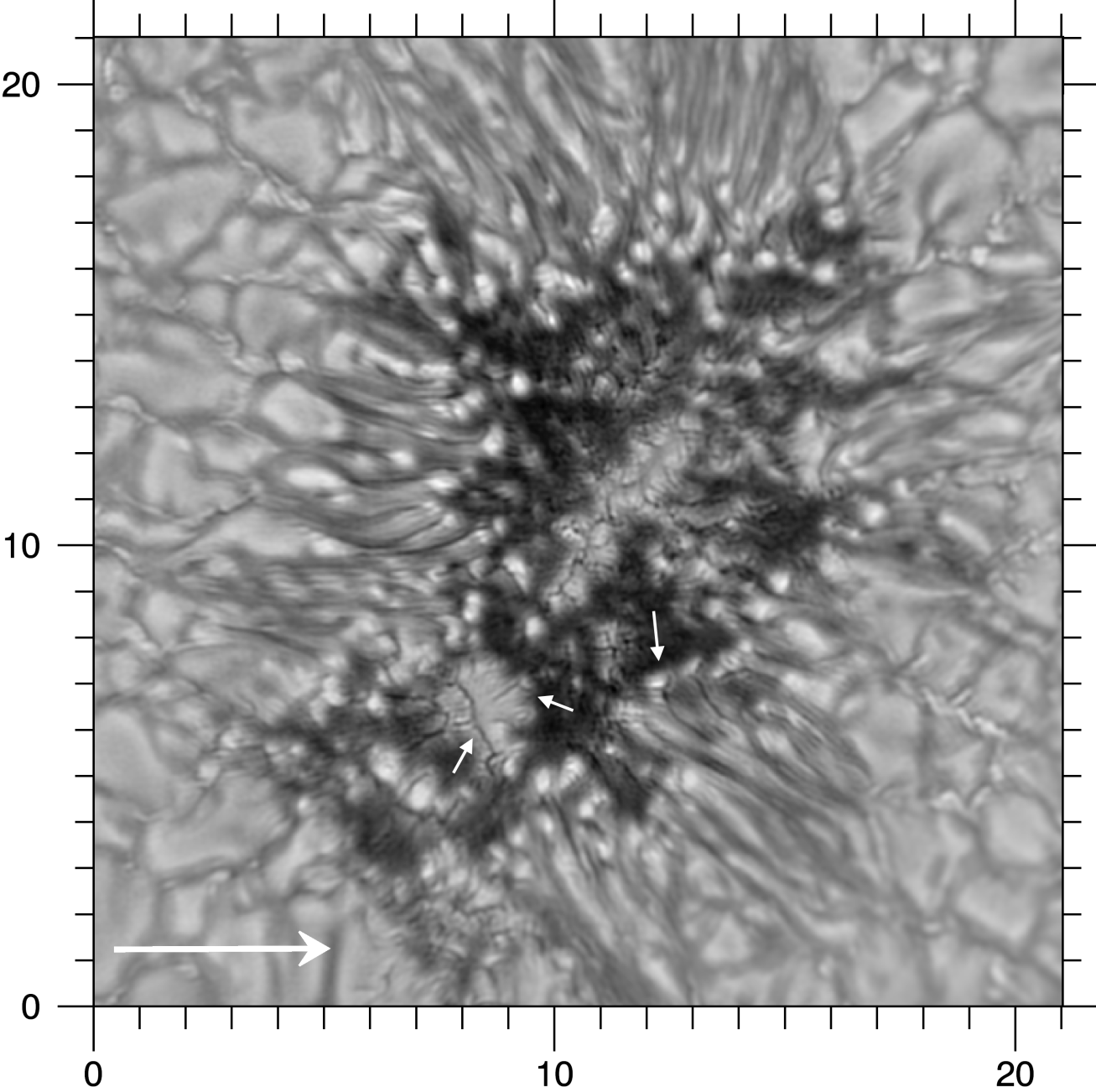}
\caption{\label{fig:3r} Spot of NOAA 11757, observed on 31 May 2013, 12:35 UT at 589\,nm. This is a subfield of Fig.~\ref{fig:1}, focusing on the spot and enhancing the umbral contrast by unsharp masking. The white arrow points at the disc centre. Tick mark units in arcsec.}
\end{figure}

\begin{figure}
   \centering
   \includegraphics*[width=8.8cm]{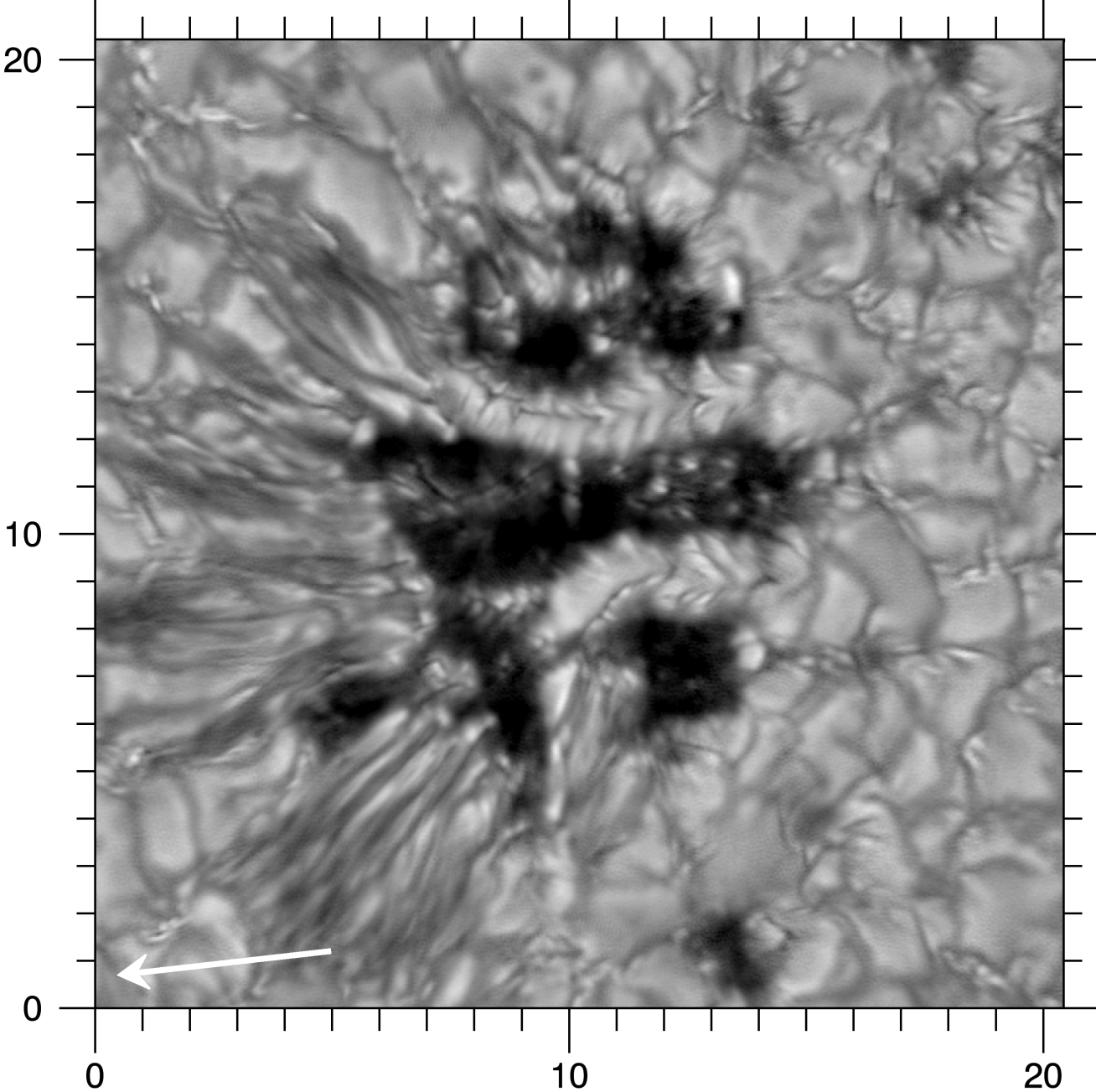} 
   \caption{Unsharp masked image of the largest sunspot of the trailing polarity of NOAA 11765 on 9 June 2013 at 09:22 UT at 486\,nm. Tick mark units in arcsec. }
   \label{fig:4r}
\end{figure}

\begin{figure}
\noindent\centering
\includegraphics*[width=9.cm]{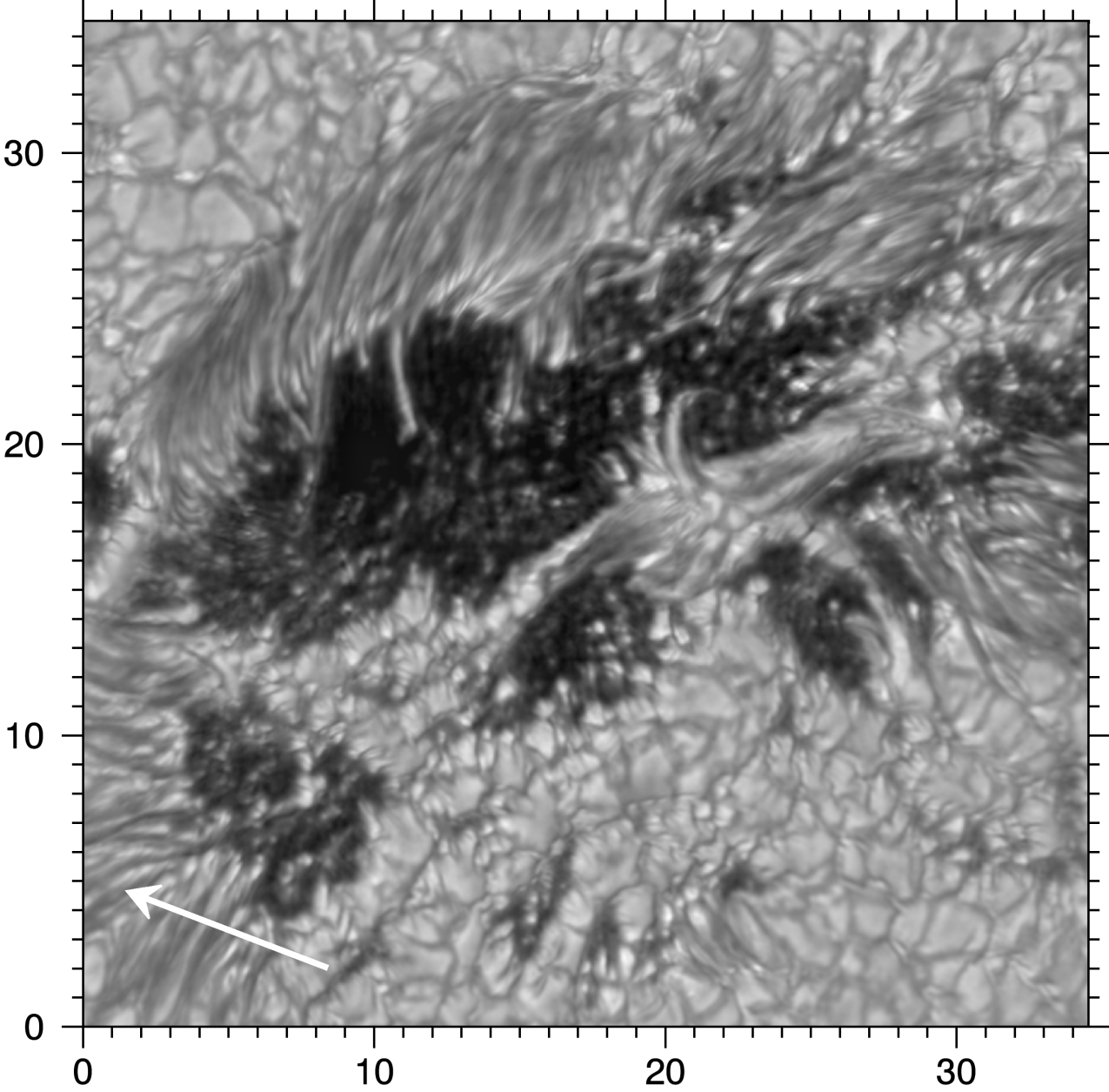}
\caption{\label{fig:2ra} Unsharp masked image of the following spot in NOAA 11768 taken on 14 June 2013 at 08:26 UT at 486\,nm. Tick mark units in arcsec. The white arrow marks the direction to disc centre.}
\end{figure}

\section{Results and discussion}

The main focus of this work is to study the morphology of the fine structure as imprints of magneto-convective processes. We selected some of the best images, zoomed into the pore or spot region and enhanced the contrast in dark regions by unsharp masking. We display these in Figs.~\ref{fig:3r}, \ref{fig:4r},  \ref{fig:2ra}, \ref{fig:2rb}, and \ref{fig:1r}. As demonstrated in the previous section, these images have a spatial resolution of up to 0\farcs08. The most prominent umbral and pore features seen in these images are umbral dots, light bridges, and dark lanes. Dark lanes are found to be ubiquitous within umbral and penumbral brightenings of all types. Dark lanes are seen along and perpendicular to light bridges in the various types of light bridges.

\subsection{Evolutionary perspective}

All four sunspots are relatively small; their umbrae show a wealth of fine structure. They exhibit various types of light bridges and umbral dots. Hence, all spots show a high degree of magneto-convective activity in the umbra. The reason presumably is that they are observed in a dynamic phase of their evolution:

(i) Figure \ref{fig:3r}: NOAA 11757 on 31 May 2013 at 12:35 UT, located at a heliocentric angle of $\theta=9^\circ$. This  small spot is isolated with no flare history. SDO/HMI continuum images\footnote{Using JHelioviewer, we viewed movies of the evolution of the active region. JHelioviewer is an open-source software by the ESA JHelioviewer team.} show that GREGOR observed it in its decaying phase of a roundish stable sunspot that rotated in on the eastern limb on 24 May and disappeared on 3 June, leaving only some plage regions. 

(ii) Figure \ref{fig:4r}:  NOAA 11765 on 9 June 2013 at 09:22 UT, located at $\theta=26^\circ$. The observed sunspot is relatively small, but is the largest spot in the trailing polarity of the AR that emerged on the disc on 5 June and disappeared on 12 June 2013. The spot was observed during its decaying phase when it was loosing part of its penumbra, and after a pore merged some 9 hours earlier. The spot changed its appearance significantly on a timescale of half a day. The merging pore created the upper umbral core. 

(iii) Figures \ref{fig:2ra}  and \ref{fig:2rb}: Two spots of opposite polarity in NOAA 11768 on 14 June 2013 at 08:26 UT and 09:26 UT, respectively. The following spot is at  $\theta=37^\circ$, the preceding at 41$^\circ$. This active region emerged on 11 June at the meridian. A first sunspot formed on 12 June; this almost completely decayed by 13 June. The pores observed on 13 June, shown in Fig.~\ref{fig:1r}, are remnants of that spot. New flux emergence in the night of 13/14 June led to the formation of two larger sunspots of opposite polarity (cf. Figs.~\ref{fig:2ra} and \ref{fig:2rb}). The sunspots showed intense activity until they rotated off the western limb on 16 June.

%

\subsection{Dark lanes and dark lane branching}

\begin{figure*}
\noindent\centering
\includegraphics*[width=16cm]{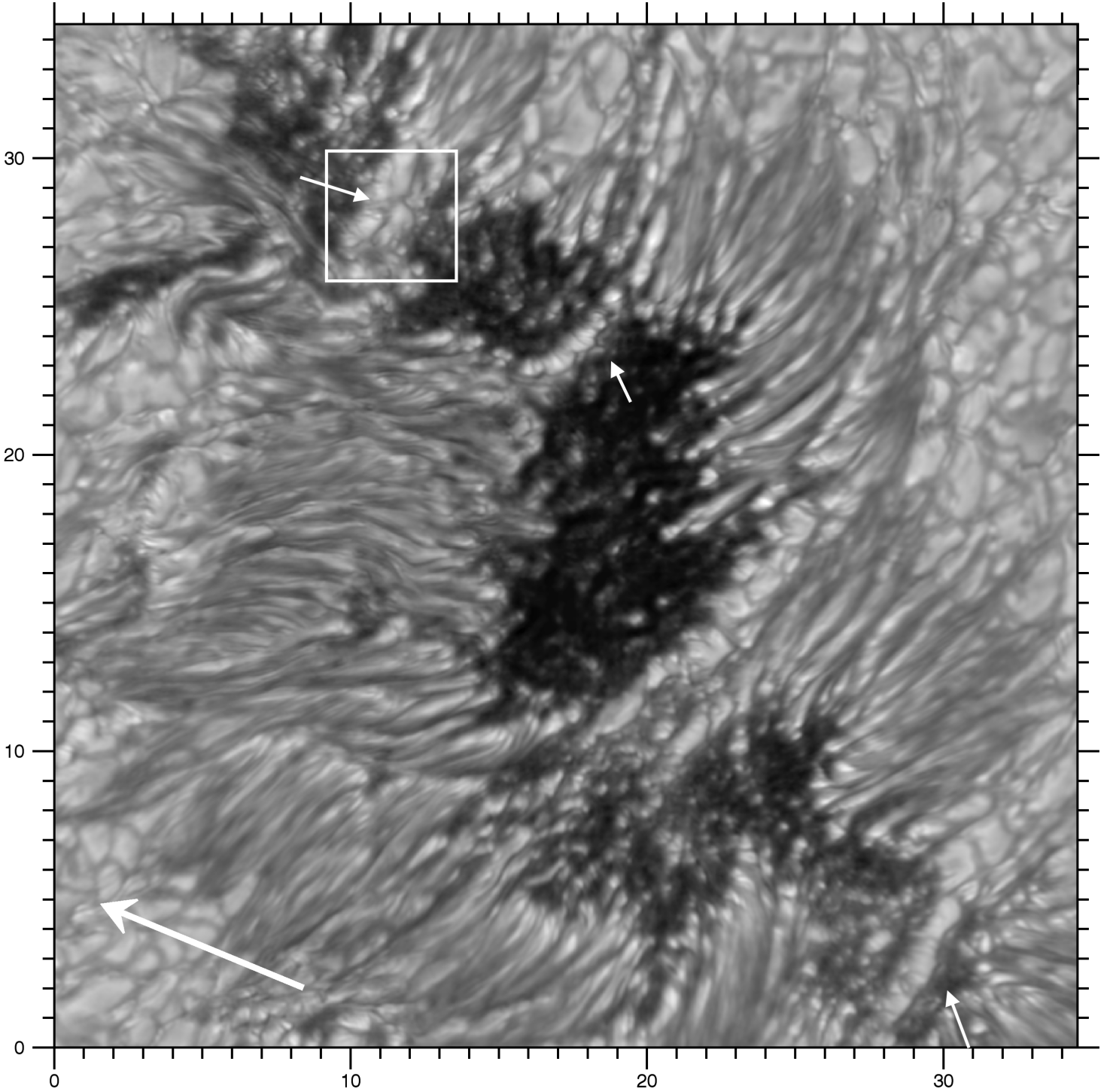}
\caption{\label{fig:2rb} Unsharp masked image of the preceding spot in NOAA 11768 taken on 14 June 2013 at 09:23 UT at 486\,nm. Tick mark units in arcsec. The large arrow marks the direction to disc centre. The small arrows indicate the three light bridges referred to in the text. The white box corresponds to the field
of view in Fig.~\ref{fig:6r}}
\end{figure*}

\begin{figure*}
\begin{center}
\includegraphics*[width=18cm]{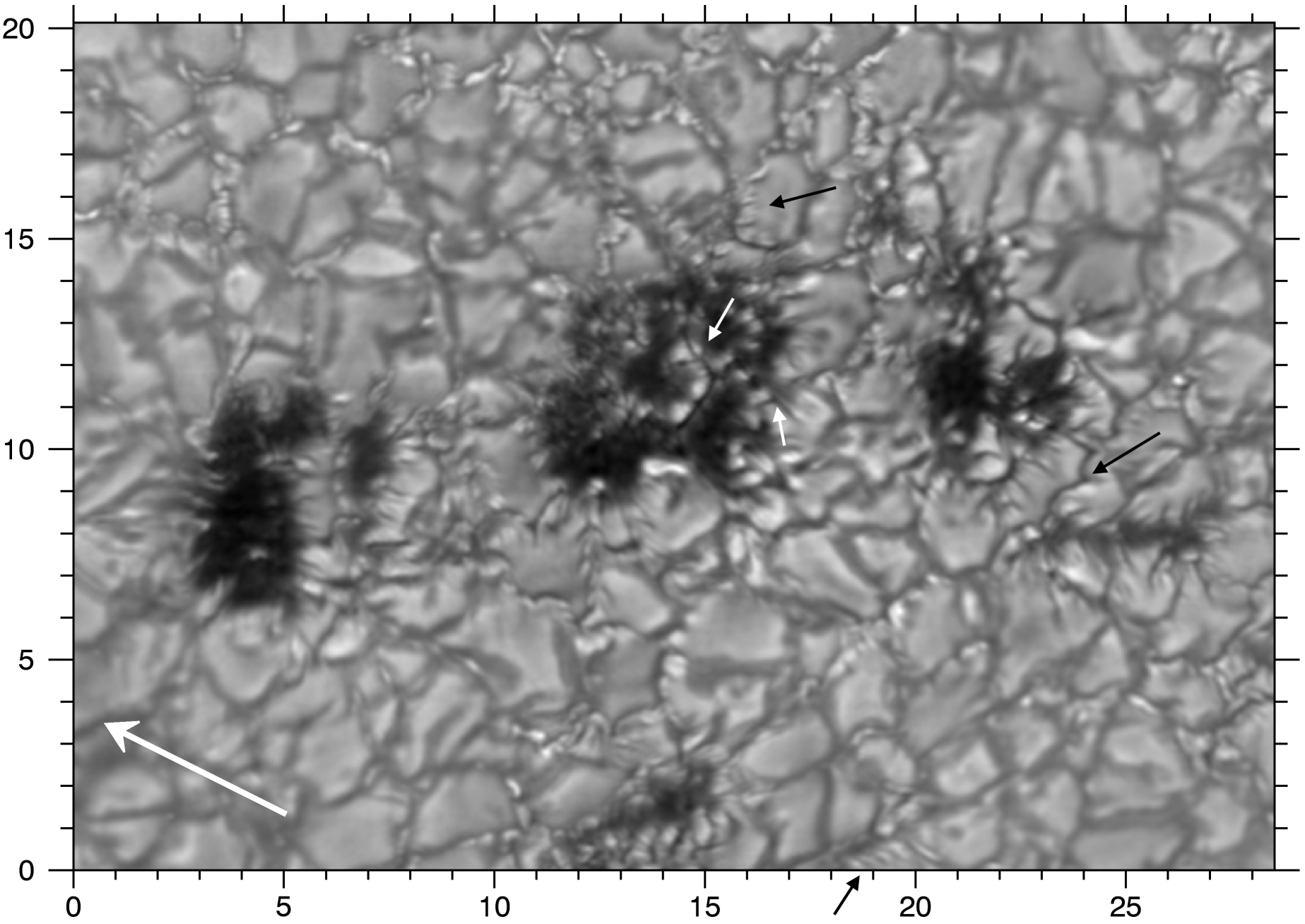}
\caption{\label{fig:1r} Active region NOAA 11768 on 13 June 2013 at 486\,nm. Tick mark units in arcsec. Dark lanes are seen as central lanes in light bridges of pores, as narrow dark intergranular lanes, as striations at the boundary of pores and granules, and as dark cores of filaments intruding into the pore. The large white arrow marks the direction to disc centre, the small arrows indicate features that are referred to in the text.}
\end{center}
\end{figure*}

\begin{figure}
\includegraphics*[width=9cm]{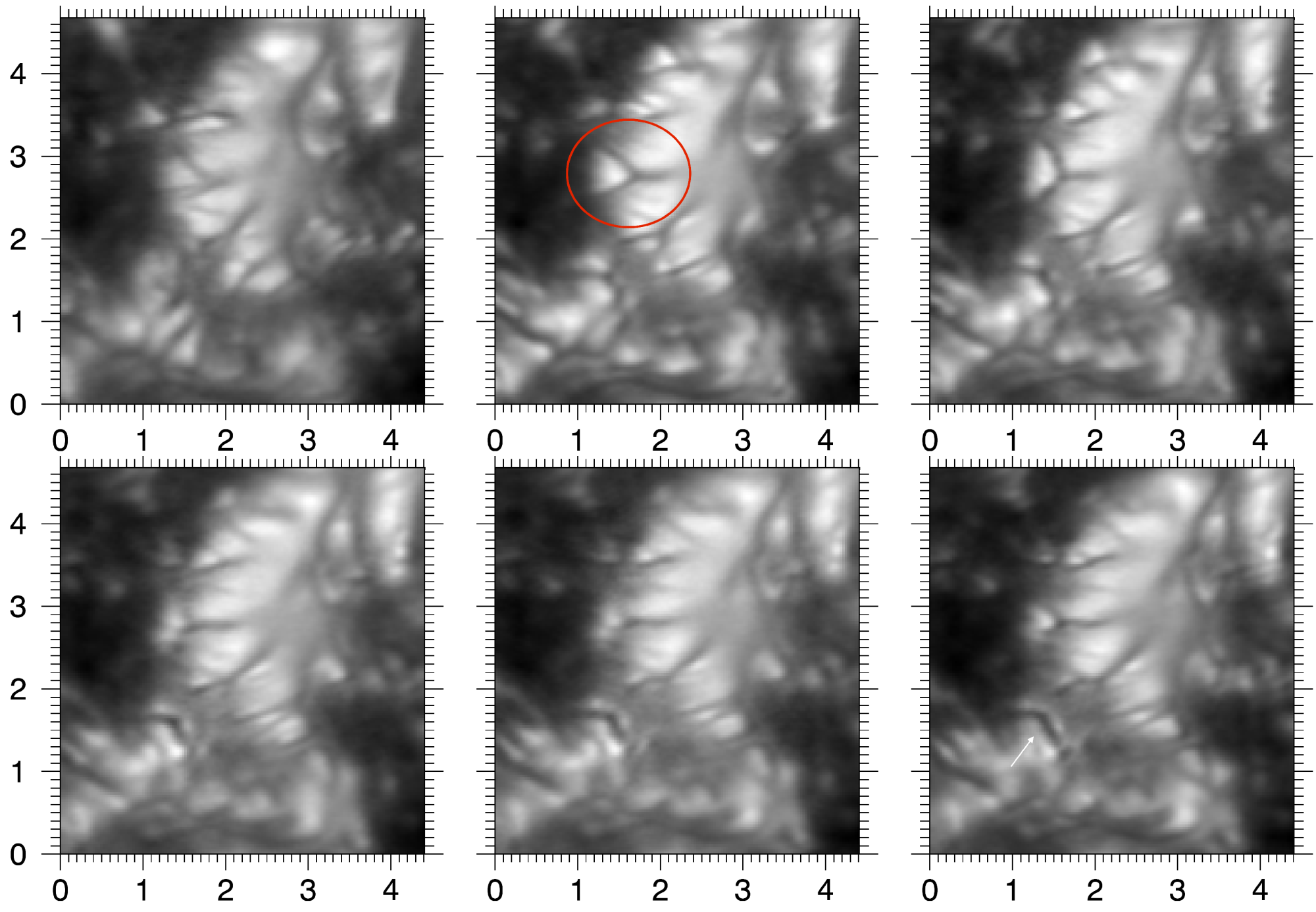}
\caption{\label{fig:6r}  Evolution (from upper left to lower right panel) of a light bridge, white box in Fig.~\ref{fig:2rb}.  Tick mark units in arcsec. The time span between the first and last image is 5\,min 20\,s: 08:58:10, 09:00:45, 09:01:50, 09:02:20, 09:03:00, and 09:03:30 UT). The time span does not include the time of the snapshot in Fig.~\ref{fig:2rb}.
}
\end{figure}

Dark lanes, or, as first described by \citet{scharmer+al2002}, `canals', are a common feature in our images.  
In Fig.~\ref{fig:1r} a central dark lane extends along the light bridge of the middle pore. At coordinate (15$\arcsec$, 12$\arcsec$) it exhibits a junction or branch. Dark lanes and junctions of dark lanes are also prominently visible in the light bridges in Fig.~\ref{fig:3r}. 
The light bridge in the mid-umbra (Fig.~\ref{fig:3r}) has a similar width as the light bridge in the pore (Fig.~\ref{fig:1r}). The light bridge in the lower left umbra (8$\arcsec$, 6$\arcsec$) of Fig.~\ref{fig:3r} is more extended ($\approx 2\arcsec$). Independently of the size of the light bridge, however, dark lanes show a characteristic width of 0\farcs1 or smaller.
Central dark lanes in most cases have branches. 

\paragraph{Central dark lanes and their branches:} Central dark lanes are also visible along the two light bridges in Fig.~\ref{fig:4r}. Lateral to the central dark lanes, dark and bright filaments are visible along the light bridge. \citet{scharmer+al2002} called
them hairs, and \citet{lites+al2004} ascribed their triangular structure to the projection effects of elevated light bridges when (i) observed at large heliocentric angles and (ii) viewed along the light bridge. The  spot in Fig.~\ref{fig:4r} is at  $\theta=26^\circ$. Viewed in a spot close to disk centre (Fig.~\ref{fig:3r}), the branches have higher contrast and are more perpendicular to the central dark lane. This indicates that the perpendicular branches in Fig.~\ref{fig:3r} and the triangular hairs in Fig.~\ref{fig:4r} are due to the same magneto-convective process, but viewed at different viewing angles. Extending the finding of Lites {\it et al.}, the following picture arises: a light bridge has the shape of a prism with the central dark lane located at its upper ridge. The lateral slopes on both sides consist of dark and bright hairs or lanes. The sides of the prism are warped such that the bright lanes are elevated with respect to the dark lanes. As a consequence, the magnetic and flow fields differ between central dark lanes and dark lane branches of light bridges. The central dark lanes would be due to increased density above the upflow as a result of piled-up matter \citep{schuessler+voegler2006}, while the lateral bright and dark lanes could correspond to hot upflow and cold downflow areas, respectively. The triangular shape is most likely produced by the expanding magnetic field of the umbra that is wrapped around the light bridge \citep[cf.][]{2006A&A...453.1079J}. Such a filamentary light bridge is sketched in the left panel of Fig.~\ref{fig:sketch} and further discussed in Sect.~\ref{subsec:new}.

In Fig.~\ref{fig:3r} the dark lanes are fairly central relative to the corresponding light bridge, with the spot being close to disc centre, $\theta=9^\circ$. In Fig~\ref{fig:2rb}, a central dark lane is visible at (18$\arcsec$, 24$\arcsec$), for example, with the spot located at $\theta=41^\circ$. The direction to Sun centre is given by the white arrow. Here, the dark lane does not split the light bridge symmetrically, but is shifted towards the limb, such that the limb side of the light bridge is almost invisible. This indicates as above \citep{lites+al2004} that the light bridge is elevated with respect to the surrounding umbra because then the central dark lane is shifted towards the limb when viewed at large heliocentric angles.

\paragraph{Y-shaped dark lanes:}
{Dark-cored bright penumbral filaments} as discovered by \citet{scharmer+al2002} are prominently visible in the spots of Figs.~\ref{fig:3r} and \ref{fig:4r}. Very often, these dark lanes show a Y shape with a separated brightening at the umbral end of the filament, for
instance at ($13\arcsec, 7\arcsec$) in Fig.~\ref{fig:3r}. This
shows that penumbral grains split (detach) from the filament. Interestingly, a similar feature is also visible in the thick light bridge at ($9\arcsec,7\arcsec$) in Fig.~\ref{fig:3r}: a Y-shaped dark lane with a brightening towards the umbra (upper right end). In contrast to branches of the central dark lane, this Y lane is unconnected to the central dark lane. The commonality with penumbral filaments could indicate that they are embedded in a highly inclined magnetic field similarly to penumbral filaments. This leads to the hypothesis that the thick light bridge is wrapped in an inclined magnetic field, such that convective instabilities can trigger the formation of filaments, as is known to occur in penumbrae. 
This filament formation could be described in terms of a rising flux tube \citep{schlichenmaier+jahn+schmidt1998a} or as an elongated magneto-convective cell \citep{rempel2012}.
Another Y-shaped dark lane is discernible at the pore boundary in Fig.\ref{fig:1r} at (16\farcs5, 11$\arcsec$), testifying that inclined fields also prevail at the pore boundary. Obviously, spectropolarimetric measurements are needed to test the first hypothesis.

A  time series of six close-ups of a light bridge  from Fig.~\ref{fig:2rb} is shown in Fig.~\ref{fig:6r}. As in Fig.~\ref{fig:3r}, a Y-shaped dark lane is visible. In the middle upper panel one Y-shaped dark lane is marked with a red circle. In the previous image (upper left panel) the dark lane has a V shape. During the subsequent 5 minutes, the V shape evolves into a Y shape and then into a simple linear shape with a bright point at the umbral end. This bright point becomes dimmer and drifts into the umbra and dissolves (not shown). In other words: a triangular granule forms within the outskirts of a thick light bridge, is squeezed towards the umbra, gradually looses its brightness, and disappears.
Similar events are also present in the same series of images.
We also observe a number of very similar events in another light bridge of the same spot and in a two-hour time series that we acquired of the spot in Fig.~\ref{fig:4r}. Therefore, these events are typical of light bridges. In animations, these events resemble a growing campfire flame that separates from the fire and fades away in the darkness. 

In the context of Fig.~\ref{fig:6r}, we note that a dark lane forms during the displayed time span and is well developed in the last frame at ($1\farcs5, 1\farcs5$). This demonstrates that the fine structure evolves on timescales of minutes.

\paragraph{Striations:}
We note that there is another type of dark lane at the border
of granules: Striations \citep{keller+al2004, spruit+al2010, carlsson+al2004}, which are bright and dark stripes that start at the granule boundary and reach into the inner granule. Examples are shown in Fig.~\ref{fig:1r} (marked with black arrows) at ($19\arcsec,0.5\arcsec$) and at ($16\arcsec,15\arcsec$). These striations are also visible in granules adjacent to the pores in Fig.~\ref{fig:1r}, although it is difficult to differentiate between striations and dark cored filaments that seem to form at the boundary of these pores. Striations were also observed along penumbral filaments \citep{ichimoto+al2007b, zakharov+etal2008, spruit+al2010} as in Fig.~\ref{fig:3r} at ($12\arcsec, 7\arcsec$). In all three cases their existence is ascribed to an inclined magnetic field that is wrapped around the granule or filament.  \citet{spruit+al2010} proposed that the interface layer between magnetic and convective plasma is unstable to the fluting instability. The corrugated surface modulates the opacity and the cooling rate, causing striations. \citet{bharti+al2012} compared observations with magnetohydrodynamics (MHD) simulations of a penumbral region to show that these dynamic striations are associated with the swaying motion of propagating kink waves.
 
\paragraph{Intergranular dark lanes:}
Intergranular lanes in quiet-Sun regions have a width of some 0\farcs3. In active regions, in the presence of magnetic flux, they may appear as thin dark lanes with a width of some 0\farcs1 and with high contrast, for example at ($24\arcsec, 9\arcsec$) in Fig.~\ref{fig:1r} \citep[see also Sect. 6.2.6 of][]{waldmann2011}. In terms of contrast and width, they are very similar to the central dark lanes in light bridges, but obviously not caused
by a density enhancement above upflows, but associated with intergranular downdrafts. Such intergranular dark lanes tend to occur next to striations or facular brightenings, indicating that they harbour a strong magnetic field that fans out with height \citep[cf. previous paragraph and][]{spruit+al2010}.

\begin{figure}
    \rotatebox{90}{\small\hspace{1.8cm}${\rm I} / {\rm I}_{\rm QS}$ \hspace{3cm} arcsec}
    \includegraphics[width=8.5cm]{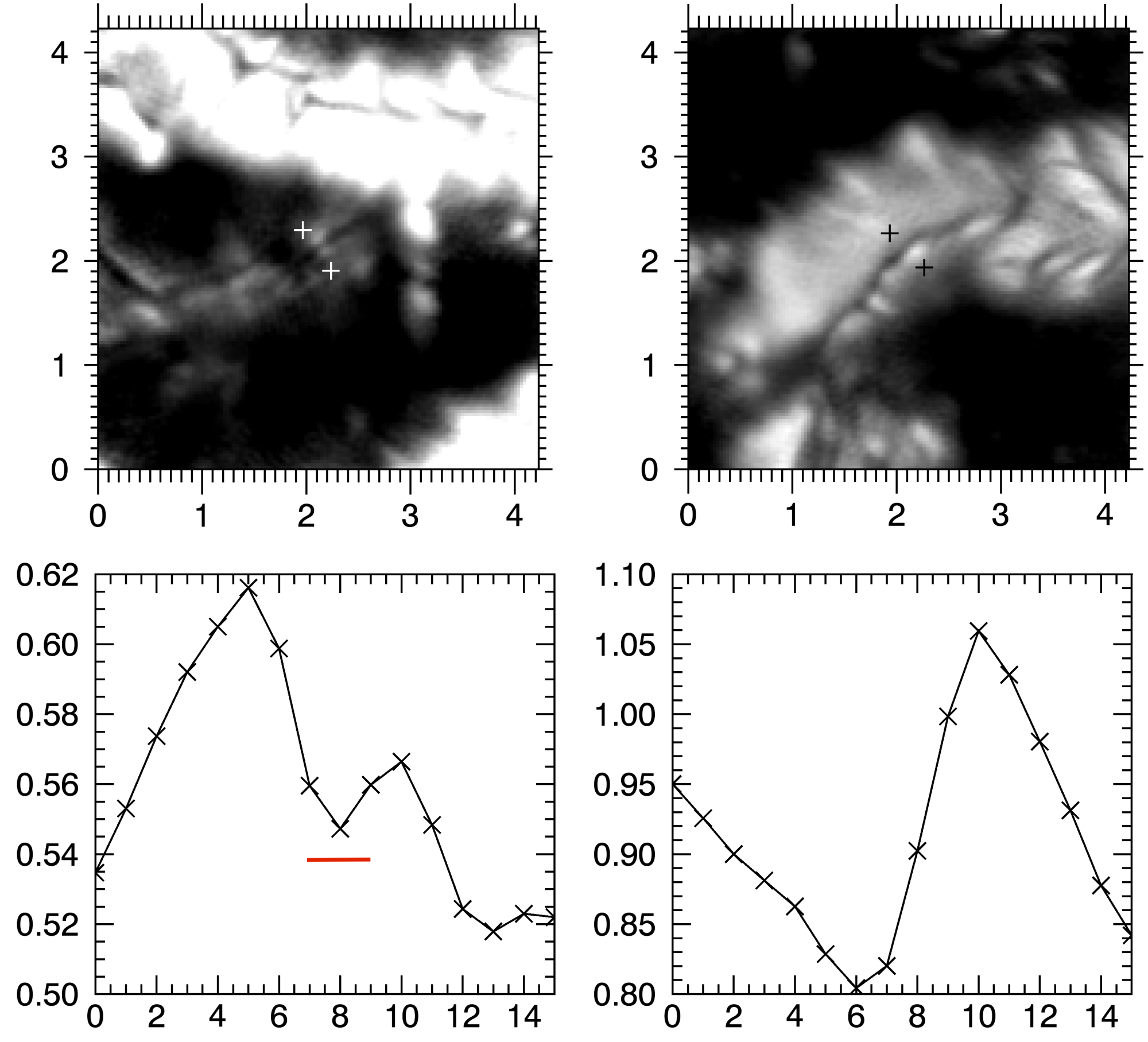} \\[-1ex]
    \rotatebox{0}{\small\hspace{1.8cm} pixels along cut\hspace{2.2cm}pixels along cut}
   \caption{Upper two panels: close-ups of two umbral regions of the spot in Fig.~\ref{fig:4r}. Tick mark units in arcsec. In the upper left panel the contrast is strongly enhanced and intensity clipped to display the faint light bridge. The two lower plots show intensity cuts from the upper left cross to the lower right cross in each image, respectively. The intensity along the cut is normalised to the averaged quiet-Sun intensity. The interpolated pixels have a distance of 0.03 arcsec. The red bar in the lower left plot marks the width of the dark lane of the faint light bridge.}
   \label{fig:7r}
\end{figure}

\subsection{New type of light bridge}\label{subsec:new}

\begin{figure}
\includegraphics[width=8.8cm]{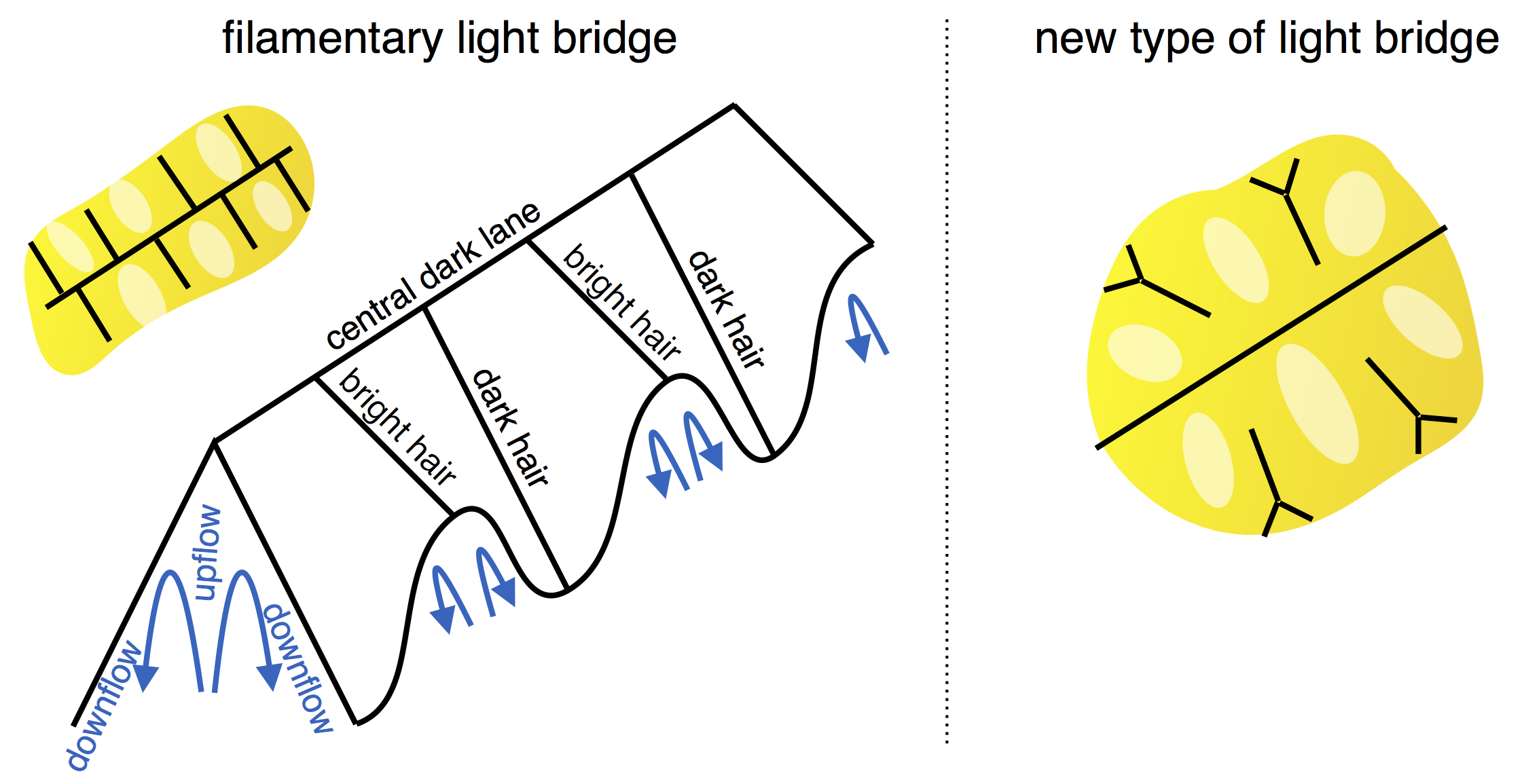}
\caption{\label{fig:sketch}{Left panel: two sketches of a filamentary light bridge. Transverse bright and dark hairs are connected to the central dark lane of the light bridge. The bright hairs resemble a granular upflow, and the dark hairs appear like intergranular lanes. Right panel: The transverse dark lanes are disconnected from the central dark lane. In many cases the transverse dark lanes are Y-shaped and appear like dark-cored penumbral filaments.
} }
\end{figure}

Light bridges with a central dark lane and branches of dark lanes or hairs are called filamentary light bridges. They have a width of some 1$\arcsec$. The light bridge in the middle of Fig.~\ref{fig:3r}, the two light bridges in Fig.~\ref{fig:4r}, the light bridge in the pore of Fig.~\ref{fig:1r}, and those discussed by \citet{lites+al2004} fall into this category. This type is sketched in the left panel of Fig.~\ref{fig:sketch}. It also shows a projected 3D view in which the bright transversal hairs are elevated relative to the dark transversal hairs.

Light bridges such as those in Figs.~\ref{fig:6r} and ~\ref{fig:3r} at ($9\arcsec,7\arcsec$) cannot be classified as granular or
filamentary, meaning that they do not fit the classification scheme of \citet{sobotka1997}. They are too thick to be called filamentary and do not resemble granulation. 
These light bridges are characterised by Y-shaped dark lanes with bright heads that are directed toward the umbra. They also have a central dark lane, but it is not shifted towards the limb when viewed at large heliocentric angles. The dark lanes are surrounded by large areas ($\varnothing  > 0.5\arcsec$) of diffuse unstructured brightness. In total, their width exceeds $\approx\!2\arcsec$. This  new type is sketched in the right panel of Fig.~\ref{fig:sketch}: the Y-shaped transverse dark lanes are unconnected to the central dark lane.

In the spot of Fig.~\ref{fig:2rb} three small arrow mark a thick light bridge at ($11\arcsec, 28\arcsec$), and two filamentary light bridges at (18$\arcsec$, 24$\arcsec$) and (30$\arcsec$, 2\farcs5). The filamentary light bridge is thought to have the shape of a prism with steep slopes towards the umbra,  which
can be explained by the projection effect. Although the thick light bridge has a similar orientation, the shift of the dark lane towards the limb is not observed. This indicates that the thick light bridge has a flat plateau. Because the disc centre side of the thick light bridge is more pronounced than the limb side, the plateau is probably elevated relative to the surrounding umbra. The light bridge plateau is a result of hot magneto-convective upflows. These bring up plasma with weak magnetic fields into the photosphere. The umbral magnetic field is wrapped around this intrusion of hot plasma and forms a canopy. At the boundary between umbra and light bridge, the field is inclined and an instability similar to the penumbra can cause the formation of filaments, leading to the Y-shaped dark lanes.

\subsection{Intensity levels of light bridges}

So far, we focused on the morphological aspect of light bridges. Now we elaborate on the intensity levels that are produced by umbral magneto-convection. In this respect, the spot in Fig.~\ref{fig:4r} is very interesting. It has two prominent light bridges, but it also shows a faint light bridge, which is located beneath the upper light bridge in the figure. We show an unsharp masked and intensity-clipped image of this faint light bridge in the upper left panel of Fig.~\ref{fig:7r}. Bright features are visible
that resemble umbral dots. Remarkably, these diffuse brightenings seem to outline a dark lane and are connected by that dark lane, or in other words, we observe a faint filamentary light bridge. It has the same morphological properties as the bright light bridge above, but at much lower intensity. The intensity cut through the faint light bridge (lower left panel) from the upper left white cross to the lower right white cross in the image reveals that the brightest regions in this faint light bridge amounts to some 62\%, normalised to the surrounding quiet Sun. The light bridge shown in the upper right panel is much brighter, and its brightest point (1.05) is even brighter than the averaged quiet Sun.

The width of the central dark lane in the faint light bridge (marked by the red bar) is only 2 pixels, that is, 0\farcs06. The width of the dark lane in the bright light bridge is not much wider, however, it corresponds to about 3 pixels, or to less than 0\farcs1.

The dark lane of the faint light bridge appears only in the presence of bright surrounding emission, and in some sections bright features are missing that would presumably outline the dark lane. Despite these gaps, the dark lane of the faint light bridge seems to be an entity over a length of more than 3$\arcsec$. In the time series of the faint light bridge with varying seeing it is present in consecutive images, supporting the statement that it is stable and extended. The associated bright features evolve on a granular timescale. Hence, in this case adjacent umbral dots are found to be connected by a dark lane. We note similarities with numerical simulation of umbral magneto-convection by \citet{schuessler+voegler2006}: the time evolution in Fig.~5 of their paper shows two adjacent faint umbral dot flow systems in the continuum images of the last three columns. Although not visible in their images, their umbral dot flow systems might be connected in the same manner as we observe it in our faint light bridge.

\subsection{Umbral magneto-convection}

The umbra exhibits a variety of morphological structures at different intensity levels. Umbral intensity on spatial averages of a few arcsec varies significantly, which is a remarkable difference to penumbrae. In the penumbra the spatial average of the transported heat always amounts to some 75\% of the quiet-Sun heat flux. This is due to a distinct difference between the umbra and the penumbra: In the penumbra the magneto-convection acts in the presence of inclined (mostly horizontal) magnetic field lines, meaning that there is one preferred lateral direction. Overturning motions manifest themselves as flows along an arch of magnetic field lines and perpendicular to the field direction by advection of magnetic field lines. 
In the umbra, the magnetic field is predominantly vertical, no lateral direction is preferred. Here the local strength of convection depends on the magnetic tension of the field, or in other words, on the magnetic field strength. This strength may vary throughout the umbra depending on evolutionary details. Although it suppresses convection, the magnetic
field in the umbra does not define a preferred direction as in the penumbra, that is, the overturning convective motions in the umbra have two lateral degrees of freedom, while they have only one degree in the penumbra. This may provide a general explanation for the more highly divers structure in the umbra. 

\subsection{Narrow penumbral filaments}

Penumbral filaments in the middle penumbra are sometimes are very long and narrow. In Fig.~\ref{fig:3r} from ($4\arcsec, 9\arcsec$) to ($7\arcsec, 9\arcsec$) narrow filaments are visible, each with a width clearly smaller than 0\farcs1, with a wavy intensity structure along their total length of some 2$\arcsec$. These filaments appear distinct from dark-cored bright filaments that are visible in the inner penumbra.

\section{Summary}

\begin{figure}
\includegraphics*[width=9cm]{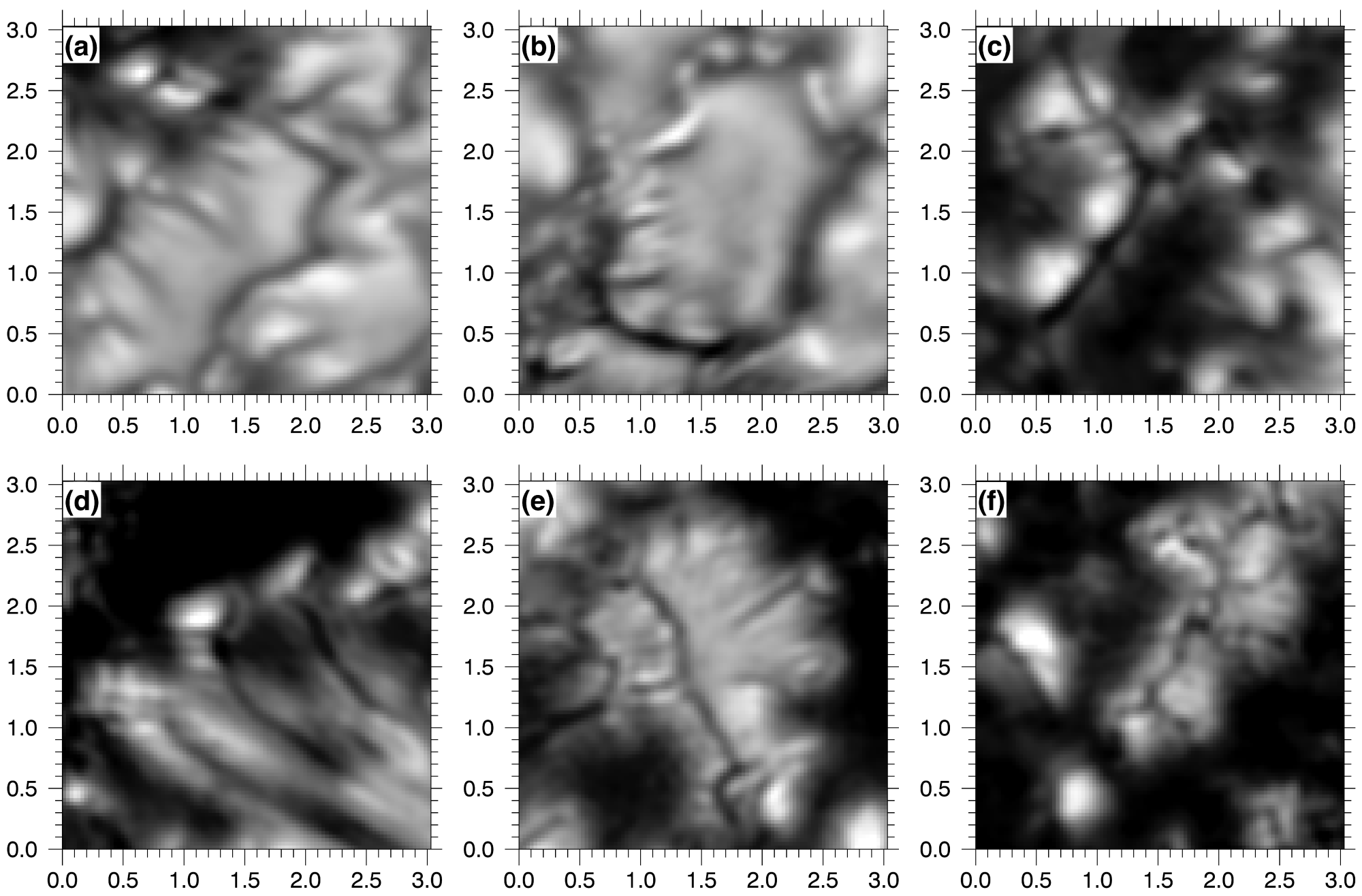}
\caption{\label{fig:paste}  Close-ups to display various types of dark lanes. All tick mark units are in arcsec. The upper three panels from Fig.~\ref{fig:1r} from left to right show (a) an
intergranular dark lane, (b) striation, and (c) a light bridge dark lane in a pore. The lower three panels from Fig.~\ref{fig:3r} from left to right show (d) a Y-shaped dark lane in a penumbral filament with striations, (e) dark lanes in a light bridge with an unconnected Y-shaped filament in the light bridge, (f) a large diffuse feature at (0\farcs9, 0\farcs4), and a dark lane system in the light bridge close to disc centre.
}
\end{figure}

We presented speckle-reconstructed images taken with the broad-band imager (BBI) at the 1.5\,m telescope GREGOR. Evaluating the characteristics of the noise, we demonstrated that the best reconstructions achieve a spatial resolution of 0\farcs08 at 589\,nm, corresponding to 60 km on the Sun.

We elaborated on the fine structure in various active regions. We observed distinct features down to the resolution limit as well as diffuse bright features on scales of 0\farcs5. Figure \ref{fig:paste} presents an overview of various features.
Describing the morphological and brightness characteristics, we reported a diversity of dark lanes: (i) central dark lanes along light bridges, (ii) branches thereof (Fig.~\ref{fig:paste} c,e, and f), (iii) dark lanes in umbral and penumbral filaments with Y shapes (Fig.~\ref{fig:paste} d and e), (iv) dark lanes as striations at the boundary of granules (Fig.~\ref{fig:paste} b), and (v) narrow intergranular dark lanes (Fig.~\ref{fig:paste} a). All of these dark lanes have a width of 0\farcs1 or less, and they are ubiquitous in high-resolution images of active regions. We discussed their appearance in the context of magneto-convective processes: central dark lanes and filamentary dark lanes might be caused by density enhancements above upflows \citep[cf.][]{schuessler+voegler2006}, branches of central dark lanes and narrow intergranular lanes might be associated with downflows, and dark lanes in striations were proposed to be caused by opacity effects that result from a fluting instability along a magnetic boundary layer \citep[][]{spruit+al2010, keller+al2004, carlsson+al2004}.

In a filamentary thin light bridge the elevated central dark lane has dark lane branches that fall off toward the umbral background and appear tilted when viewed along the light bridge at large heliocentric angles. In thick light bridges, a central dark lane and branches also exist, but the branches are unconnected
and are Y-shaped with a detached bright grain towards the umbra. The Y-shaped dark lanes in light bridges look very similar to dark-cored penumbral filaments intruding into the umbra. Their inner brightening is separated by a Y-shaped dark lane. They evolve on timescales of minutes. The inner bright grain moves towards the umbra and fades away. We ascribed the similarity
in their nature to penumbral filaments to the inclined fields that are wrapped around the light bridges. 

Light bridges occur on different intensity levels. We observed one faint light bridge with a maximum intensity of 60\% of the quiet-Sun average. It also featured a dark lane: The dark lane is outlined by adjacent bright features. It has gaps, but appears as an entity over a length of more than $3\arcsec$.

The umbral fine structure is diverse. It is more diverse in brightness than in the penumbra. We ascribe this difference to the strongly inclined magnetic field in the penumbra. This causes it to show a preferred lateral direction. In the vertical magnetic field of the umbra, overturning convection can occur in both lateral directions. Here the mode and effectivity of the convective heat transport depends on the strength of the magnetic field, which varies throughout the umbra. Hence various modes of magneto-convection create a broad diversity of umbral structures.
We recall that convective heat transport is required to explain the observed brightness of the umbra and that magneto-convective signatures are expected to be ubiquitous in the umbra, even in the darkest umbral regions. It will be interesting to investigate
whether dark lanes are also present in the faintest umbral features.

We discussed the observed active region fine structure in terms of magneto-convective processes, but spectro-polarimetric measurements are clearly needed at a spatial resolution better than 0\farcs1 to infer the velocity and magnetic fields. These are then to be compared with theoretical models and considerations. 


\begin{acknowledgements}
We thank Christoph Kuckein, Reza Rezaei, Morten Franz, and the anonymous referee for most valuable comments on the manuscript.
The 1.5 meter GREGOR solar telescope was built by a German consortium under the leadership of the Kiepenheuer-Institut f\"ur Sonnenphysik in Freiburg with the Leibniz-Institut f\"ur Astrophysik Potsdam, the Institut f\"ur Astrophysik G\"ottingen, and the Max-Planck-Institut f\"ur Sonnensystemforschung in G\"ottingen as partners, and with contributions by the Instituto de Astrof{\'i}sica de Canarias and the Astronomical Institute of the Academy of Sciences of the Czech Republic. This work was supported by the SOLARNET project (www.solarnet-east.eu), funded by the European Commission's FP7 Capacities Programme under the Grant Agreement 312495.

\end{acknowledgements}

\bibliographystyle{aa}

\end{document}